%% file: draft.tex
\documentclass[epj,iicol,numbook]{svjour}
\input{settings.tex}
\usepackage{cuted}

\usepackage[square,numbers,sort&compress]{natbib}

\begin{document}
\title{Einstein and Jordan frame correspondence in quantum cosmology: Expansion-collapse duality}

\author{Dipayan Mukherjee\thanks{dipayanmkh@gmail.com} \and Harkirat Singh Sahota\thanks{harkirat221@gmail.com}}

\institute{\emph{Indian Institute of Science Education and Research Mohali}\\
  Sector 81, SAS Nagar, Manauli PO 140306, Punjab, India}

\date{Received: date / Revised version: date}

\abstract{
  The conformal correspondence between FLRW universes in the Einstein and Jordan
  frames allows for an expansion-collapse duality -- an always expanding
  Einstein frame universe can have a dual Jordan frame description that is
  contracting forever. The scenario eventually runs into an apparent paradox.
  When the contracting Jordan frame universe becomes sufficiently small, the
  classical description becomes inadequate and the universe is expected to
  develop quantum characteristics. However, at this time, the corresponding
  Einstein frame universe is expected to behave classically, due to the
  arbitrarily large size it has grown to. The conformal map here appears to be
  providing a duality between a quantum effect-dominated universe and a
  classical universe. We investigate the status of the conformal map at the
  quantum level in such a scenario, focusing on addressing this paradox. The
  Einstein and Jordan frame universes are quantized using the Wheeler-DeWitt
  prescription. We show that the classical conformal map holds at the quantum
  level when compared through expectation values of scale factors. The relative
  quantum fluctuation in the scale factor becomes conformally invariant, it
  increases in both the past and future directions according to the internal
  clock. Expectedly, the quantum fluctuations in the collapsing Jordan frame
  increase as it shrinks towards singularity. More surprisingly, the quantum
  fluctuations in the expanding Einstein frame increase as well, even as its
  classical scale factor becomes larger. Despite having drastically different
  cosmological evolutions, the rise in quantum characteristics in a collapsing
  frame implies the same in its expanding counterpart, thereby resolving the
  apparent paradox.
  \PACS{{PACS-key}{discribing text of that key} \and {PACS-key}{discribing text of that key} } 
} 
\titlerunning{Einstein and Jordan frame correspondence in quantum cosmology:
  Expansion-collapse duality} \authorrunning{Mukherjee \& Sahota}

\maketitle
\section{Introduction}
It is well-known that scalar-tensor theories of gravity can be put in the form
of Einstein's gravity with a minimally coupled scalar-field in a conformally
connected spacetime, where the former and latter representations are referred to
as the \emph{Jordan} and \emph{Einstein frames},
respectively~\cite{faraoni04,fujii.maeda03}. The two conformally connected
frames are mathematically equivalent by construction; however, whether the
frames can be considered as `physically equivalent' is highly
debated~\cite{faraoni.gunzig99, postma.volponi14,catena.ea07, artymowski2013,
  faraoni.gunzig.ea98,faraoni.ea07, flanagan2004, bahamonde.odintsov.ea16,
  chiba.ea13, briscese.elizalde.ea07, banerjee2016, kamenshchik2015,
  racioppi2022, capozziello.nojiri.ea06*1, rondeau2017}.
While the conformal transformation removes the non-minimal coupling between the
scalar field and curvature in the Jordan frame action, this comes at a cost. An
ordinary matter component in the Jordan frame becomes non-trivially coupled to
the scalar field in the Einstein frame, resulting in an extra `fifth force' term
in the geodesic equation of a material particle. The modification in the
geodesic equation can be seen as a violation of the \emph{Einstein's equivalence
  principle}. This is often used as the main argument for the Einstein frame not
being the physical frame. The counter arguments to this is largely based the
notion of `running units' in the Einstein
frame~\cite{faraoni.ea07,deruelle2011,dicke1962,prokopec.ea13}. As originally
suggested by Dicke~\cite{dicke1962}, the conformal transformation of the metric
naturally leads to spacetime dependent scaling of the units of fundamental
physical quantities, such as length, time, and mass. It is argued that, if one
consistently adopts this running units in the Einstein frame, then the two
conformally connected frames can be regarded as physically equivalent
(see~\cite{faraoni.ea07} for a review).

Regardless of Dicke's argument, the Einstein and Jordan frame descriptions
generally differ in their cosmological evolutions~\cite{felice.ea10,
  nojiri.odintsov.ea17, nojiri.odintsov11, bahamonde.odintsov.ea17, fertig.ea16,
  ijjas.ea15, wetterich14a, mukherjee.jassal.ea21, mukherjee2022}. Although
different conformal frames are mathematically equivalent, the equations of
motion in these frames may lead to quite different properties of the scale
factors, Hubble parameters, accelerations, and any higher-order derivatives of
the scale factors in the two frames. One may use this duality to study standard
cosmological models of the physical universe through conformally connected
universes with contrasting features. Such conformal maps have interesting
implications. For example, in~\cite{bahamonde.odintsov.ea17}, it is shown that
an accelerated expansion of space can alternatively be realized as a
decelerating phase in a conformally connected universe, demonstrating that
cosmic acceleration is a frame dependent effect. In~\cite{wetterich14a}, the
author introduces a `slow freeze' model of the early universe as a dual
description of the hot big-bang model, where the universe is contracting instead
of expanding in the radiation and matter-dominated eras. Consequently, the
universe is free from a big-bang-like singularity in the slow-freeze
description. In~\cite{ijjas.ea15}, the authors propose an early-time `anamorphic
phase' of the universe, during which the cosmological evolution shows features
of both the inflationary model and the ekpyrotic model depending on the choice
of different conformally invariant criteria. By combining features of both the
inflationary and ekpyrotic scenarios, the anamorphic universe model attempts to
bypass some shortcomings associated with inflationary and bouncing models when
considered individually. The `conflation model', introduced
in~\cite{fertig.ea16}, also explores the duality between the inflationary and
the ekpyrotic model. During the `conflation'-phase, the accelerated expansion in
one frame maps to a contracting phase in another frame.

It is also possible to realize the dark energy-driven standard late-time
cosmological models in conformal frames with contrasting cosmological
evolutions. In~\cite{mukherjee.jassal.ea21}, a class of quintessence models is
shown to be conformally dual to $f(R)$ gravity driven Jordan frames, where the
Jordan frame may undergo a collapse. In~\cite{mukherjee2022}, it is shown that
the cosmological evolutions of the $\Lambda$CDM and quintessence models in the
Einstein frames always correspond to bouncing or collapsing Jordan frames driven
by Brans-Dicke theories. With suitable choice for the Brans-Dicke parameter, the
current epoch of the standard cosmological evolution can be aligned with a
bounce in the Jordan frame, during which the Jordan frame scale factor becomes
arbitrarily small.

In general, the expansion-collapse duality between the Einstein and Jordan
frames has interesting consequences. Given the choice of the theories in the two
frames, an ever-expanding universe in the Einstein frame may correspond to a
Jordan frame universe which is contracting indefinitely. As the collapsing
Jordan frame universe becomes sufficiently small, one may expect it to develop
quantum characteristics; for example, the quantum uncertainties in different
cosmological observables may increase as the scale factor gets smaller. At the
same time, the scale factor in the dual Einstein frame universe keeps on
increasing. This raises the questions -- what is the status of the classical
conformal map when one considers the quantum descriptions of the Einstein and
Jordan frame universes in such a scenario? Subsequently, how do the increasing
quantum uncertainties in the collapsing Jordan frame universe correspond to the
uncertainties in the expanding Einstein frame? One can speculate that the
quantum variances in the Einstein frame may get suppressed due to the expansion
of space. If this is true, then the conformal correspondence seems to provide a
map between a quantum-effect dominated universe and a universe with negligible
quantum corrections. Alternatively, if the quantum fluctuations in the Einstein
and Jordan frames evolve similarly despite their contrasting cosmological
evolutions, then this would indicate that the increase in quantum
characteristics is a frame-independent effect.

Previous studies have explored different aspects of the Einstein-Jordan frame
duality in quantum descriptions~\cite{ashtekar2003, grumiller2002, nojiri2001,
  grumiller2003, elizalde1994, fujii1990, flanagan2004, artymowski2013,
  kamenshchik2015, banerjee2016, almeida2018, ohta2018, faraoni.ea07}. For
example, in \cite{artymowski2013}, the Einstein and Jordan frame representations
of the Brans-Dicke theory are studied in the loop quantum cosmology framework.
The loop quantizations in different frames are shown to be leading to
non-equivalent results. In~\cite{banerjee2016}, a Brans-Dicke model with a
perfect fluid is quantized in the Wheeler-DeWitt prescription in both frames. In
this case, the Einstein and Jordan frame wave packet solutions are found to be
incompatible when compared via the conformal map. However, it is shown in a
subsequent study~\cite{pandey.ea17} that the Einstein and Jordan frame wave
packets do become conformally connected in the absence of additional matter
components.

In this paper, we investigate the expansion-collapse duality between the
conformal frames in a fully quantum mechanical framework according to the
Wheeler-DeWitt quantization scheme~\cite{dewitt1967,kiefer2012}. We consider the
simple example of the Brans-Dicke model without a potential in the Jordan frame
that corresponds to a massless canonical scalar field in the Einstein frame. In
the classical description, one may choose a Brans-Dicke parameter for which the
Jordan frame contracts indefinitely, corresponding to an ever-expanding Einstein
frame. To describe the expansion-collapse duality at the quantum level, the
Einstein and Jordan frame universes are individually quantized using the
Wheeler-DeWitt prescription, following~\cite{pandey.ea17}. We derive the
expectation values for relevant cosmological quantities in both frames. To see
how the classical conformal correspondence translates into the quantum
description, we find the relations between different Einstein and Jordan frame
quantities through their expectation values and compare these relations with
their classical counterparts. In this regard, our approach differs from that
of~\cite{pandey.ea17}, where the status of the conformal map is investigated at
the level of wave packets in the two frames. The maps between the Einstein and
Jordan frame expectation values also determine whether an expansion-collapse
duality between the frames is possible in the quantum description. Finally, we
see how quantum fluctuations of different quantities are related via the
conformal map and address whether the increasing nature of quantum
characteristics in one frame implies the same in the other frame.

The paper is organized as follows. In~\cref{sec:class-expans-coll}, we discuss
the classical expansion-collapse duality between a Jordan frame with the
Brans-Dicke model, and its corresponding Einstein frame.
In~\cref{sec:canon-descr-conf}, we set up the classical Hamiltonian formalism in
the Einstein and Jordan frames. The universes in the two frames are then
quantized in~\cref{sec:quant-descr-conf}, and we obtain expectation values and
quantum fluctuations of different cosmological quantities.
In~\cref{sec:conf-map-expans}, we investigate the status of the conformal map at
the quantum level and discuss how the quantum fluctuations in the two frames are
related. We end with a summary and discussion in~\cref{sec:summary-conclusion}.

\section{Classical expansion-collapse duality}
\label{sec:class-expans-coll}
Depending on the scalar field model, the cosmological evolution in the Einstein
and Jordan frame descriptions may be drastically different. In this paper, we
are interested in a scenario where an ever-expanding Einstein frame universe
corresponds to a Jordan frame that is contracting indefinitely. In fact, one can
realize such an expansion-collapse duality even through a simple Brans-Dicke
model in the Jordan frame. In this section, we consider the original Brans-Dicke
model with zero potential in the Jordan frame and show that the
expansion-collapse duality can be achieved for certain choice of the Brans-Dicke
parameter.

\begin{figure}
  \centering
  \includegraphics[width=\columnwidth]{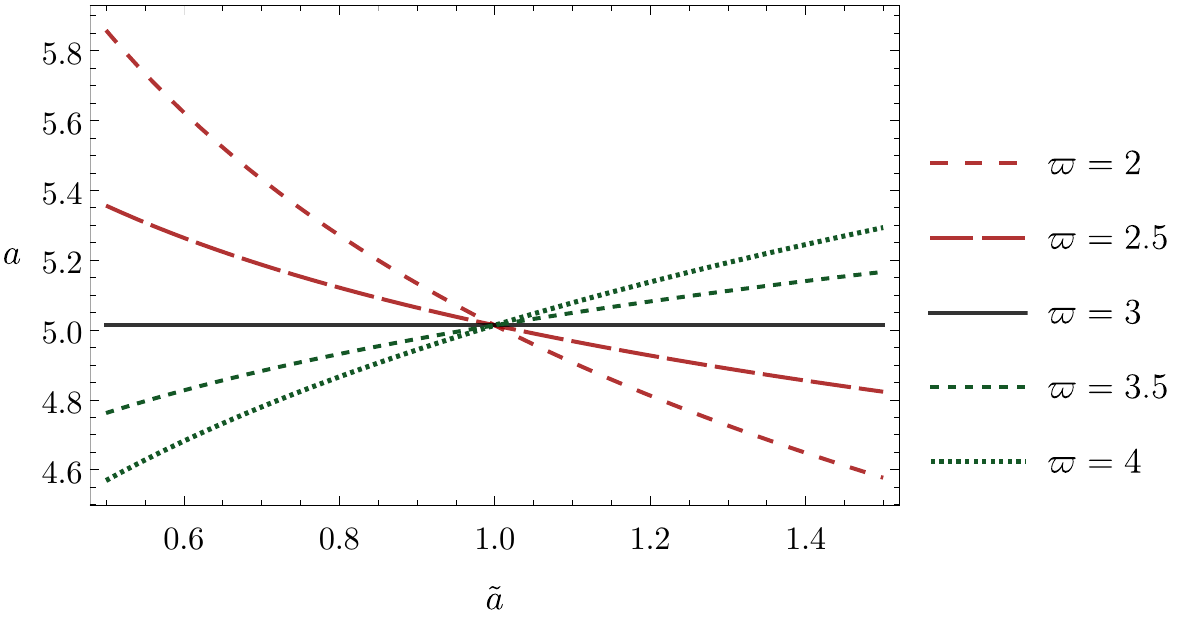}
  \caption{Expansion-collapse duality in the classical description. The plots
    show evolution of the Jordan frame scale factors $(a)$ corresponding to
    different Brans-Dicke models specified by the parameter $\varpi$, with
    respect to the Einstein frame scale factor $(\tilde{a})$ (where
    $\lambda_0=1$). For the models with $\varpi>3$, the Jordan frame universes
    expand with the Einstein frame universe; while for $\varpi<3$, the Jordan
    frames are always collapsing. $\varpi=3$ leads to a static Jordan frame
    universe. }
  \label{fig:1}
\end{figure}

Let us consider the Brans-Dicke theory action in the Jordan frame, governed by
the Brans-Dicke scalar field $\lambda$ and the metric $g_{ab}$,
\begin{align}
  \label{eq:1} 
  S_{J}^{BD} &= \int \d^4 x \sqrt{-g} \left( \frac{\lambda}{16 \pi} R - \frac{w_{\text{BD}}}{16 \pi \lambda} g^{ab} \p_a \lambda \p_b \lambda  \right),
\end{align}
where $R$ is the Ricci scalar in the Jordan frame and $\wbd$ is the constant
Brans-Dicke parameter~\cite{faraoni04,fujii.maeda03,brans2014}. The conformal
transformation
\begin{align}
  \label{eq:2}
  \tilde{g}_{ab} = G \lambda g_{ab}
\end{align}
leads to the Einstein frame action
\begin{align}
  \label{eq:3}
  S_{E} =\int \d^4 x \sqrt{-\tilde{g}} \left[\frac{1}{2 \kappa^2} \tilde{R} -  \frac{1}{2}  \tilde{g}^{ab} \p_a\varphi \p_b\varphi - V(\varphi) \right],
\end{align}
where $G$ is the Newtonian constant of gravity, $\kappa^2 = 8 \pi G$,
$\tilde{R}$ is the Ricci scalar for the metric $\tilde{g}_{ab}$, and the
Einstein frame scalar field $\varphi$ is identified as~\cite{faraoni04}
\begin{align}
    \label{eq:4}
  \diff{\varphi}{\lambda} = \sqrt{\frac{\varpi}{2}} \frac{1}{\kappa \lambda},
\end{align}
where we define the parameter
\begin{align}
 \label{eq:5} 
  \varpi \equiv 2 \wbd + 3.
\end{align}
Note that $\varpi>0$ is required in order for the field $\varphi$ to be real. We
consider that both the Jordan and Einstein frames have spatially flat FLRW
metrics,
\begin{subequations}
  \label{eq:6}
  \begin{align}
    \label{eq:7}
    g_{ab} &\equiv \text{diag} \left[ -1, a^2(t), a^2(t), a^2(t) \right],\\
    \tilde{g}_{ab} &\equiv \text{diag} \left[ -1, \at^2(\tilde{t}), \at^2(\tilde{t}), \at^2(\tilde{t}) \right],
  \end{align}
\end{subequations}
where the Einstein and Jordan frame scale factors $(\at, a)$ and coordinate
times $(\tilde{t}, t)$ are then related via the conformal transformation as
\begin{subequations}
  \label{eq:8}
  \begin{align}
    \label{eq:9}
    \at &= \sqrt{G \lambda}a,\\
    \label{eq:10}
    \dif \tilde{t} &= \sqrt{G \lambda} \dif t .
  \end{align}
\end{subequations}
The energy density, pressure and equation of state parameter of the homogeneous
Einstein frame scalar field $\varphi$ are given by the usual definitions,
\begin{align}
  \label{eq:11}
  \rho_\varphi =  \frac{1}{2} \left(\diff{\varphi}{\tilde{t}}\right)^2,\quad P_\varphi =  \frac{1}{2} \left(\diff{\varphi}{\tilde{t}}\right)^2,\quad
  w_\varphi = \frac{P_\varphi}{\rho_\varphi} = 1.
\end{align}
The possibility of an expansion-collapse duality between the Einstein and Jordan
frames can easily be observed by expressing the Jordan frame scale factor $(a)$
in terms of the Einstein frame scale factor $(\tilde{a})$ from~\cref{eq:9}. To
obtain the relation, we rewrite~\cref{eq:11} as
\begin{subequations}
  \label{eq:12}
  \begin{align}
    \label{eq:13}
    \left( \diff{\varphi}{\te}  \right)^2 &= \rho_\varphi + P_\varphi = \rho_\varphi(1 + w_\varphi),\\
    \label{eq:14}
    \diff{\varphi}{\at} &=  \frac{1}{\at \tilde{H}} \sqrt{\rho_\varphi(1 + w_\varphi)},
  \end{align}
\end{subequations}
where the Einstein frame Hubble parameter $\tilde{H}$ is constrained via the
Friedmann equation
\begin{align}
  \tilde{H}^2 \equiv \left( \frac{1}{\tilde{a}} \diff{\tilde{a}}{\tilde{t}} \right)^2 = \frac{\kappa^2}{3}  \rho_\varphi.
\end{align}
Choosing an expanding Einstein frame $(\tilde{H}>0)$ and replacing $\tilde{H}$
in~\cref{eq:14}, the Einstein frame scalar field can be solved as
\begin{align}
  \label{eq:15}
  \kappa (\varphi - \varphi_0) &= \sqrt{6} \ln \tilde{a},
\end{align}
where $\varphi_0 = \varphi(\tilde{a} = 1)$, and we have put $w_\varphi=1$
from~\cref{eq:11}. For simplicity, we set $\varphi_0=0$, therefore fixing the
origin of the Einstein frame scalar field as $\varphi(\at=1)=0$. The Brans-Dicke
scalar field $\lambda$ can be written in terms of $\varphi$, from~\cref{eq:4},
\begin{align}
  \label{eq:16}
  \lambda(\varphi) = \lambda_0 \exp \left( \sqrt{\frac{2}{\varpi}} \kappa \varphi \right),
\end{align}
where $\lambda_0 = \lambda(\varphi=0)$. Finally, from~\cref{eq:9}, the Jordan
frame scale factor can be written in terms of the Einstein frame scale factor as,
\begin{align}
\label{eq:17}
    a = \sqrt{\frac{8 \pi}{\kappa^2 \lambda_0}} \at^{1 - \sqrt{\frac{3}{\varpi}}},
\end{align}
where we have used~\cref{eq:16,eq:15}. It is evident that the condition required
for an increasing Einstein frame scale factor to map to a decreasing Jordan
frame scale factor is given by
\begin{align}
  \label{eq:18}
  \diff{a}{\tilde{a}} < 0 \iff \varpi < 3.
\end{align}
Noting that the Einstein frame scale factor $\at$ monotonically increases with
the Einstein frame coordinate time $\tilde{t}$, and the Jordan frame coordinate
time $t$ is monotonically changing with $\tilde{t}$ (see~\cref{eq:10}), the
Jordan frame scale factor $a(t)$ is always decreasing if $\varpi<3$. Therefore,
\emph{for an expanding Einstein frame universe, the conformally dual Jordan frame
universe contracts (expands) indefinitely if it is driven by a Brans-Dicke model
with $\varpi<3$ ($\varpi>3$)} (see~\cref{fig:1}).

The expansion-collapse duality also becomes evident if one considers the
relation between the Jordan and Einstein frame Hubble parameters, $H$ and
$\tilde{H}$. One can obtain from~\cref{eq:9,eq:10}
\begin{align}
  \label{eq:19}
  H \equiv \frac{1}{a} \diff{a}{t} = \sqrt{\frac{\kappa^2 \lambda}{8 \pi}} \left( 1 - \sqrt{\frac{3}{\varpi}} \right) \tilde{H},
\end{align}
therefore, given an expanding Einstein frame $(\tilde{H}>0)$, the Jordan frame
Hubble parameter $H$ becomes negative for $\varpi<3$, depicting a collapsing
Jordan frame.

From~\cref{eq:17} we see that depending on the value of the integration constant
$\lambda_0$, $a$ can become arbitrarily small for a large $\at$. After a time
when the Jordan frame scale factor becomes sufficiently small, the quantum
effects in the Jordan frame can no longer be ignored. One may assume that the
quantum effects become more robust as the Jordan frame contracts further. At the
same time, the conformally dual Einstein frame scale factor keeps on increasing
and one can anticipate that the quantum effects are no longer relevant. We are
interested in how the quantum effects in the contracting Jordan frame maps to
that of the expanding Einstein frame universe. In particular, we seek whether
the quantum features in the Einstein frame are suppressed as the space becomes
arbitrarily large, or whether the Einstein frame quantum effects increase
similarly to its Jordan frame counterpart, despite the expansion of space.

In order to address this question, we seek whether there exists a conformal map
between the Einstein and Jordan frames at the quantum level. Then we find how
the quantum fluctuations of different cosmological quantities in the two frames
are related via the conformal map. We begin with treating both the Einstein and
Jordan frame universes in the minisuperspace formalism of quantum cosmology. The
following section introduces Hamiltonian formalism in the Einstein and Jordan
frame descriptions. Later on, this is used the quantize the conformally
connected universes.

\section{Canonical descriptions of the conformal frames}
\label{sec:canon-descr-conf}
In this section, we introduce Hamiltonian formalism in both Einstein and Jordan
frame universes, where the Jordan frame is driven by a Brans-Dicke model with
zero potential, and the corresponding Einstein frame consists of a minimally
coupled canonical massless scalar field.

\subsection{Hamiltonian formalism in the Einstein frame}
Let us consider the line element in the Einstein frame to be
\begin{align}
  \d \tilde{s}^2 = -\tilde{n}^2(\tilde{t}) \d \tilde{t}^2 + \tilde{a}^2(\tilde{t}) \delta_{\alpha \beta} \d x^\alpha \d x^\beta,
\end{align}
where $\tilde{n}$ is the lapse function in the Einstein frame. Using this in the Lagrangian of the Einstein frame
\begin{align}
  \label{eq:20}
  \tilde{L}=\sqrt{-\tilde{g}} \left[\frac{1}{2 \kappa^2} \tilde{R} -  \frac{1}{2}  \tilde{g}^{ab} \p_a\varphi \p_b\varphi  \right], 
\end{align}
and adding appropriate GHY (Gibbons-Hawking-York) boundary term to make the 
variational principle well-posed, we obtain
\begin{align}
  \label{eq:21}
  \tilde{L} &= - \frac{3 \at \dot \at^2}{\kappa^2 \tilde{n}} + \frac{\at^3 \dot \varphi^2}{2 \tilde{n}}.
\end{align}
Note that in this subsection, the overdots denote derivatives with respect to
$\tilde{t}$. Under the variable transformation
\begin{align}
  \label{eq:22}
  \tilde{\chi} = \ln \at,
\end{align}
the Lagrangian takes the form
\begin{align}
  \label{eq:23}
  \tilde{L} = \frac{1}{\tilde{n}} \exp \left( 3 \tilde{\chi} \right)  \left( - \frac{3}{\kappa^2} \dot{\tilde{\chi}}^2
  + \frac{1}{2} \dot \varphi^2\right).
\end{align}
The primary constraint of the system is $P_{\tilde{n}}\approx 0$, where $P_j$ are
the conjugate momenta associated with the variable $j$. Legendre transformation
of this Lagrangian system leads to the Einstein frame Hamiltonian
\begin{align}
  \tilde{\H} &= \tilde{n} \exp \left(-3 \tilde{\chi}  \right) \left( -\frac{\kappa^2}{12} P_{\tilde{\chi}}^2 + \frac{1}{2} P_{\varphi}^2 \right).
\end{align}
Following~\cite{pandey.ea17}, we do a canonical transformation in the scalar
field sector
\begin{align}
  \Te &= \frac{\varphi}{P_{\varphi}}, \\
  P_{\Te} &= \frac{1}{2} P_{\varphi}^2,
\end{align}
leading to the Hamiltonian that is linear in the momentum conjugate to the
variable $\Te$,
\begin{align}
  \tilde{\H} \left( \tilde{\chi}, \tilde{n}, \Te, P_{\tilde{\chi}}, P_ {\tilde{n}}, P_{\Te}\right) &= \frac{\tilde{n}}{12} \e^{-3 \tilde{\chi}} \left( - P_{\tilde{\chi}}^2 + 12 P_{\Te} \right),
\end{align}
here we have set $\kappa^2=1$. The consistency equation for the primary
constraint, $\dot{P}_{\tilde{n}}\approx 0$ leads to the secondary (Hamiltonian)
constraint,
\begin{align}
	 \label{eq:24}
	\tilde{\mathcal{H}} &\equiv - P_{\tilde{\chi}}^2 + 12 P_{\tilde{T}} \approx 0.
\end{align}
The Hamilton's equations of motion in the Einstein
frame become
\begin{subequations}
  \label{eq:25}
  \begin{align}
    \label{eq:26}
    \dot {\tilde{T}} &= \tilde{n} \exp \left( -3 \tilde{\chi} \right), \\
    P_{\tilde{T}} &\equiv P_{\tilde{T}} (\text{constant}),\\
    \label{eq:27}
    \dot {\tilde{\chi}} &= -\frac{\tilde{n}}{6} \exp \left( - 3 \tilde{\chi} \right) P_{\tilde{\chi}}, \\
    P_{\tilde{\chi}} &\equiv P_{\tilde{\chi}} (\text{constant}).
  \end{align}
\end{subequations}
Using~\cref{eq:26,eq:27}, one can solve for the classical trajectory
$\tilde{\chi}(\tilde{T})$ as
\begin{align}
	\label{eq:28}
	\tilde{\chi}(\tilde{T}) &= - \frac{1}{6} P_{\tilde{\chi}} \tilde{T} + c_1,\\
	\label{eq:29}
	\tilde{\chi}(\tilde{T}) &= \tilde{H}_0 \tilde{T},
\end{align}
where in the second line we have identified
$\tilde{H}_0 = \tilde{H}(\tilde{\chi}=0) = - P_{\tilde{\chi}}/6$
from~\cref{eq:30} and set $c_1=0$, thus choosing the origin of $\tilde{T}$ as
$\tilde{\chi}(\tilde{T}=0)=0$.

For the choice of $\tilde{n}=1$, that is, taking $\tilde{t}$ to be the comoving
time coordinate in the Einstein frame, the Hubble parameter in this frame can be written
in terms of the phase-space variables as
\begin{subequations}
  \label{eq:30}
  \begin{align}
    \label{eq:31}
    \tilde{H} &= \frac{1}{\at} \diff{\at}{\tilde{t}} = \dot{\chi},\\
    \label{eq:32}
    \tilde{H} &= - \frac{1}{6} P_{\tilde{\chi}} \exp \left( - 3 \tilde{\chi} \right). 
  \end{align}
\end{subequations}

\subsection{Hamiltonian formalism in the Jordan frame}
Let us consider the Jordan frame line element
\begin{align}
  \label{eq:33}
  \d s^2 = - n^2(t) \d t^2 + a^2(t) \delta_{\alpha \beta} \d x^\alpha \d x^\beta,
\end{align}
where $n$ is the Jordan frame lapse function. Using this in the Lagrangian of
the Jordan frame,
\begin{align}
  \label{eq:34}
  L = \frac{\sqrt{-g}}{16 \pi} \left(\lambda R - \frac{\wbd}{\lambda} g^{a b} \p_a \lambda \p_b \lambda \right),
\end{align}
and adding appropriate boundary term we get
\begin{align}
  L = \frac{1}{16 \pi n} \left(\wbd a^3 \frac{\dot{\lambda}^2}{\lambda} - 6 \lambda
  a \dot{a}^2 - 6 a^2 \dot{\lambda} \dot{a} \right).
\end{align}
Here the overdots denote derivatives with respect to $t$.
Following~\cite{pandey.ea17}, we choose the variable transformation
\begin{subequations}
  \label{eq:35}
  \begin{align}
    \label{eq:36}
    a &= \exp \left( - \frac{\alpha}{2} + \beta \right),\\
    \label{eq:37}
    \lambda &= \exp \left( \alpha \right),
  \end{align}
\end{subequations}
leading to the Lagrangian
\begin{align}
  L = \frac{1}{16 \pi n} \exp \left( - \frac{\alpha}{2} + 3 \beta \right) \left[\frac{1}{2} \varpi \dot{\alpha}^2 - 6 \dot{\beta}^2  \right].
\end{align}
Again, $P_{n}\approx 0$ is the primary constraint on the phase-space of this
system, and the Jordan frame Hamiltonian can be obtained as
\begin{subequations}
  \begin{align}
    \H &= \frac{16 \pi n}{24} \exp \left( \frac{\alpha}{2} -  3 \beta \right) \left( -P_\beta^2 + \frac{12}{\varpi} P_\alpha^2 \right).
  \end{align}
\end{subequations}
The canonical transformation
\begin{subequations}
  \label{eq:38}
\begin{align}
  \label{eq:39}
  T &= \frac{\alpha}{P_\alpha},\\
  \label{eq:40}
  P_T &= \frac{P_\alpha^2}{2},
\end{align}
\end{subequations}
is used to then replace the phase-space variables
$(\alpha, P_\alpha) \to (T, P_T)$, leading to the Hamiltonian
\begin{align}
  \label{eq:41}
  &\H \left(n, \beta, T, P_n, P_\beta, P_T\right) \nonumber\\
 &=\frac{16 \pi n}{24} \exp \left( \frac{T \sqrt{P_T}}{\sqrt{2}} - 3 \beta \right) \left( -P_\beta^2 + \frac{24}{\varpi} P_T \right).
\end{align}
The consistency equation of the primary constraint leads to the Hamiltonian constraint
\begin{align}
	 \label{eq:42}
	\mathcal{H} &\equiv -P_\beta^2 + \frac{24}{\varpi} P_T \approx 0.
\end{align}
The Hamilton equations of motion in the Jordan frame take the forms
\begin{subequations}
  \begin{align}
    P_{\beta} &\equiv P_\beta \text{(constant)},\\
    \label{eq:43}
    \dot \beta &= \frac{16 \pi n}{24} \exp \left(\frac{T \sqrt{P_T}}{\sqrt{2}} - 3 \beta \right) \left( -2 P_{\beta}\right)\\ 
    P_T &\equiv P_T \text{(constant)},\\
    \label{eq:44}
    \dot T &\approx  \frac{16 \pi n}{24} \exp \left(\frac{T \sqrt{P_T}}{\sqrt{2}} - 3 \beta \right) \left( \frac{24}{\varpi} \right).
  \end{align}
\end{subequations}
One can solve for the classical trajectory in the Jordan frame
from~\cref{eq:43,eq:44} as
\begin{subequations}
  \label{eq:45}
  \begin{align}
    \beta &= -\frac{\varpi P_{\beta}}{12} T + c_2,
  \end{align}
\end{subequations}
where $c_2$ is the integration constant. Using this and the variable
transformations~\cref{eq:35,eq:38,eq:42} we find
\begin{align}
 \label{eq:46} 
  \chi  &= \ln a = -\frac{\alpha}{2}+\beta \nonumber \\
        &=  \sqrt{\frac{P_T}{6}} \left(  \sqrt{\varpi} - \sqrt{3} \right)T - \sqrt{\frac{\varpi}{12}} \ln \lambda_0 - \ln \sqrt{\frac{\kappa^2}{8 \pi}},
\end{align}
where we have set
$c_2 = - \sqrt{\varpi/12} \ln \lambda_0 - \ln (\kappa^2/(8 \pi))$ in order to be
consistent with the conformal transformation in \cref{eq:17}.

The Hubble parameter in the Jordan frame can be written as a function of the
phase-space variables as (see \cref{eq:39,eq:40})
\begin{subequations}
  \begin{align}
    \label{eq:47}
    H &= \dot \chi = \diff{}{t} \left( \beta - \sqrt{\frac{P_T}{2}} T\right),\\
    \label{eq:48}
    H &= 16 \pi  \exp \left( T \sqrt{\frac{P_T}{2}} - 3 \beta \right) \left( - \frac{P_\beta}{12} - \frac{1}{\varpi} \sqrt{\frac{P_T}{2}} \right);
  \end{align}
\end{subequations}
here we have chosen $n=1$, therefore setting $t$ to be the comoving time in the
Jordan frame.

\section{Quantum descriptions of the conformal frames}
\label{sec:quant-descr-conf}
With the phase-space structure at hand, one can proceed to quantize the systems
in both frames. According to Dirac's prescription, the primary constraints imply
that the wave functions are independent of the lapse choices; while, the
secondary constraints lead to the Wheeler-DeWitt equations in both frames. In
this analysis, we have deparameterized the systems such that the scalar field
$\varphi$ serves as a clock in the Einstein frame, whereas, the clock in the
Jordan frame is provided by $\alpha$, which is a combination of both the
Brans-Dicke field and the scale factor.

\subsection{Quantum description in the Einstein frame}
In the coordinate representation, the Hamiltonian constraint,~\cref{eq:24},
leads to the Wheeler-DeWitt equation
$\widehat{\tilde{\mathcal{H}}} \tilde{\psi} = 0$,
\begin{align}
  \label{eq:49}
  \left( \diffp[2]{}{{\tilde \chi}} - 12 i \diffp{}{{\tilde T}} \right) \tilde{\psi} &= 0.
\end{align}
The operator is self-adjoint with the Hilbert space $L^2(\mathbb{R},\d \tilde\chi)$. The Wheeler-DeWitt equation has the form of a free particle Schr\"odinger equation and it admits a plane wave solution,
\begin{align}
  \tilde{\psi}_m \left( \tilde{\chi}, \tilde{T} \right) = \exp \left( i \left( m^2 \tilde{T} -  \sqrt{12} m \tilde{\chi} \right) \right).
\end{align}
We construct a normalized wave packet from the stationary states using a
Gaussian distribution
\begin{align}
  \label{eq:50}
  \tilde{\phi}(m) = \exp \left( - \tilde{\gamma} (m - \tilde{m}_0)^2 \right),
\end{align}
  \begin{align}
    \label{eq:51}
    \tilde{\Psi} \left( \tilde{\chi}, \tilde{T} \right) &=  \int_{-\infty}^{\infty} \d m\ \tilde{\phi}(m) \tilde{\psi}_m \left( \tilde{\chi}, \tilde{T} \right) \nonumber \\
    &= \left( \frac{\pi^{3/2}}{\sqrt{6 \tilde{\gamma}}} \right)^{-1/2} \sqrt{\frac{\pi}{\tilde{\gamma} - i \tilde T}} \nonumber \\
    &\times \exp \left( -\frac{3 \tilde{\chi}^2 - i \tilde{m}_0 \gamma \left( \tilde{m}_0 \tilde{T} - 2 \sqrt{3} \tilde{\chi}  \right)}{\tilde{\gamma} - i \tilde{T}} \right).
  \end{align}
The wave packet constructed here satisfies DeWitt's criteria as
$\lim\limits_{\tilde{\chi}\rightarrow-\infty}\tilde{\Psi}(\tilde{\chi},\tilde{T})=0$
and hints at singularity resolution~\cite{kiefer2012,kiefer2019}. The
expectation of $\tilde{\chi}$ can be obtained as
\begin{align}
  \label{eq:52}
  \langle \tilde{\chi} \rangle_E = \int_{- \infty}^{\infty} \d \chi \  \tilde{\Psi}^* \tilde{\chi} \tilde{\Psi} = \frac{\tilde{m}_{0}}{\sqrt{3}} \tilde{T},
\end{align}
where the subscript $E$ denotes that the expectation is evaluated with respect
to the Einstein frame wave packet. If one chooses the wave packet parameter as
\begin{align}
  \label{eq:53}
  \tilde{m}_0 &=  \sqrt{3} \tilde{H}_0 = \sqrt{\frac{2}{\varpi}} \sqrt{P_T},
\end{align}
where the second equality follows from the conformal transformation, then
\begin{align}
  \label{eq:54}
  \langle\tilde\chi\rangle_E &= \tilde{H}_0\tilde{T}.
\end{align}
Thus, $\Braket{\tilde{\chi}}_E$ always follows the classical
trajectory,~\cref{eq:29}, and thereby suggesting singularity non-resolution for
the state that satisfy DeWitt's criteria.

The variance of $\tilde{\chi}$ is found to be
\begin{align}
  \label{eq:55}
  \Delta_E \tilde{\chi} = \langle \tilde{\chi}^2\rangle_E - \langle \tilde{\chi}\rangle_E^2 = \frac{\tilde{T}^2 + \tilde{\gamma}^2}{12 \tilde{\gamma}}.
\end{align}
We see that the variance becomes minimum as $\abs{\tilde{T}} \to 0$, and
diverges as $\abs{\tilde{T}} \to \infty$. The expectation value of the Einstein
frame scale factor is obtained as
\begin{subequations}
  \label{eq:56}
  \begin{align}
    \label{eq:57}
    \langle \at \rangle_E &= \Braket {\exp \left( \tilde{\chi} \right)}_E = \exp \left( \frac{\tilde{m}_0}{\sqrt{3}} \tilde{T} + \frac{1}{24} \left( \frac{\tilde{T}^2}{\tilde{\gamma}} + \tilde{\gamma} \right)\right),
                           \intertext{which can be put in the form}
                           \label{eq:58}
                           \langle \at \rangle_E & =  \exp \left( \langle\tilde{\chi}\rangle_E + \frac{\Delta_E\tilde{\chi}}{2}\right) = \tilde{a}(\tilde{T}) \exp \left( \frac{\Delta_E\tilde{\chi}}{2}\right),
  \end{align}
\end{subequations}
where $\tilde{a}(\tilde{T})$ is the classical trajectory (from~\cref{eq:29}),
\begin{align}
  \label{eq:59}
  \tilde{a} (\tilde{T}) = \exp \left( \tilde{H}_0 \tilde{T} \right),
\end{align}
and we have used the parameter choice in~\cref{eq:53}. As we can see,
$\Braket{\at}_E$ does not follow the classical trajectory for all $\tilde{T}$,
in fact, the deviation from the classical trajectory is dictated by the
fluctuation in $\tilde{\chi}$ and it increases as $\abs{\tilde{T}}$ becomes
larger. The expectation value of the scale factor suggests a regular universe with a non-zero minimum at $\tilde{T}=-12\tilde{H}_0\tilde{\gamma}$ and therefore singularity resolution.

The variance and the relative fluctuation in the scale factor can be obtained as
\begin{subequations}
  \begin{align}
    \label{eq:60}
    \Delta_E \at &= \langle \at^2 \rangle_E - \langle \at \rangle_E^2 \nonumber \\
                &= \Braket{\exp \left( 2 \tilde{\chi}  \right)}_E - \Braket{\exp \left( \tilde{\chi} \right)}_E^2,\\
    \Delta_E \at &= \left( \exp \left( \frac{\tilde T^2 + \tilde{\gamma}^2}{12 \tilde{\gamma}} \right) - 1 \right) \nonumber \\
                & \times \exp \left( \frac{2 \tilde{m}_0}{\sqrt{3}} \tilde{T} + \frac{\tilde{T}^2}{12 \tilde{\gamma}} + \frac{\tilde{\gamma}}{12} \right),\\
    \frac{\sqrt{\Delta_E \at}}{\Braket{\at}_E} &= \sqrt{\exp \left( \frac{\tilde T^2 + \tilde{\gamma}^2}{12 \tilde{\gamma}} \right) - 1}=\sqrt{\exp \left( \Delta_E\tilde{\chi} \right) - 1}.
  \end{align}
\end{subequations}
The relative fluctuation $\sqrt{\Delta_E \at}/\Braket{\at}_E$ is at its minimum
when $\abs{\tilde{T}} \to 0$ and it diverges as $\abs{\tilde{T}} \to \infty$.
One can construct the operator for the Einstein frame Hubble
parameter,~\cref{eq:32}, as
\begin{align}
  \label{eq:61}
  \hat {\tilde{H}} = - \frac{1}{12} \left( \exp \left( - 3 \hat{\tilde{\chi}} \right) \hat{P}_{{\tilde{\chi}}} + \hat{P}_{{\tilde{\chi}}} \exp \left( - 3 \hat{\tilde{\chi}} \right) \right).
\end{align}
With the parameter choice from~\cref{eq:53}, the expectation value of
$\hat{\tilde{H}}$ can be written as
\begin{subequations}
  \label{eq:62}
  \begin{align}
    \label{eq:63}
  \Braketil{\tilde{H}}_E &= \tilde{H}_0 \exp \left( - 3 \tilde{H}_0 \tilde{T} \right)  \nonumber \\
 &\times \exp \left( \frac{3 \left(\tilde{T}^2 + \tilde{\gamma}^2\right)}{8 \tilde{\gamma}} \right) \left( 1 - \frac{\tilde{T}}{4 \tilde{\gamma}\tilde{H}_0} \right),\\
  &= \tilde{H} (\tilde{T}) \exp\left(\frac{9}{2}\Delta_E\tilde{\chi}\right) \left( 1 - \frac{\tilde{T}}{4 \tilde{\gamma}\tilde{H}_0}\right)
\end{align}
\end{subequations}
where $\tilde{H}(\tilde{T})$ is the classical expression,
\begin{align}
  \label{eq:64}
  \tilde{H} = \tilde{H}_0 \exp \left( - 3 \tilde{H}_0 \tilde{T} \right).
\end{align}
The deviation of $\Braketil{\tilde{H}}$ from its classical trajectory increases
as $\abs{T}$ gets larger. The expectation value of the Hubble parameter also suggests a bouncing universe with the bounce occurring at $\tilde{T}=4\tilde{H}_0\tilde{\gamma}$.

\subsection{Quantum description in the Jordan frame}
The Hamiltonian constraint in the Jordan frame,~\cref{eq:42}, leads to the
Wheeler-DeWitt equation $\widehat{\mathcal{H}} \psi = 0$,
\begin{align}
  \label{eq:65}
  \left( \diffp[2]{}{{ \beta}} - \frac{24}{\varpi} i \diffp{}{{T}} \right) \psi = 0
\end{align}
The differential operator is self-adjoint on the Hilbert space
$L^2(\mathbb{R},\d \beta)$. As in the Einstein frame, the Jordan frame
Wheeler-DeWitt equation has the form of a free particle Schr\"odinger equation
and it admits a plane wave solution,
\begin{align}
  \psi_m \left(  {\beta},  {T} \right) = \exp \left( i \left( m^2  {T} -  \sqrt{\frac{24}{\varpi}} m  {\beta} \right) \right).
\end{align}
We construct a normalized wave packet from the stationary state using a Gaussian
distribution
\begin{align}
  \label{eq:66}
  \phi(m) = \exp \left( - \gamma (m - m_0)^2 + i \beta_0 m\right),
\end{align}
  \begin{align}
    \label{eq:67}
    &\Psi \left(  {\beta},  {T} \right) =\int_{-\infty}^{\infty} \d m\  \phi(m)   \psi_m \left(  {\beta},  {T} \right) \nonumber \\
    &= \left( \frac{\pi^{3/2}}{2\sqrt{3}} \sqrt{\frac{\varpi}{\gamma}} \right)^{-1/2} \sqrt{\frac{\pi}{\gamma - i  T}} \nonumber \\
    &\times \exp \left( -\frac{\left( \sqrt{\frac{6}{\varpi}} \beta - \frac{\beta_0}{2}  \right)^2 - i m_0 \gamma \left( m_0  {T} - 2 \sqrt{\frac{6}{\varpi}} \beta + \beta_0 \right)}{\gamma - i  {T}} \right).
  \end{align}
In this case as well, the DeWitt's criteria is satisfied by the wave packet,
$\lim\limits_{\beta\rightarrow-\infty}\Psi(\chi,T)=0$, hinting at singularity
resolution. The expectation value of $\chi$ with respect to the Jordan frame
wave packet can be obtained as
\begin{subequations}
\begin{align}
  \label{eq:68}
  \langle \chi \rangle_J &= \int_{- \infty}^{\infty}  \d \beta \  \Psi^*  \left(\beta - \sqrt{\frac{P_T}{2}} T \right)  \Psi  \\
  &= \left(\frac{m_0 \sqrt{\varpi}}{\sqrt{6}}-\frac{\sqrt{P_T}}{\sqrt{2}} \right) T + \sqrt{\frac{\varpi}{6}} \frac{\beta_0}{2}.
\end{align}
\end{subequations}
For the choice of parameters
\begin{subequations}
  \label{eq:69}
  \begin{align}
    \label{eq:70}
    m_0 &= \sqrt{P_T},\\
    \beta_0 &= - 2 \sqrt{\frac{6}{\varpi}} \ln \left( \frac{1}{\sqrt{8 \pi}} \lambda_0^{\sqrt{\frac{\varpi}{12}}} \right),
  \end{align}
\end{subequations}
$\Braket{\chi}_J$ follows the classical trajectory,~\cref{eq:46}, for all $T$,
\begin{align}
  \langle \chi \rangle_J =  \sqrt{\frac{P_T}{6}} \left( \sqrt{\varpi} - \sqrt{3} \right) T + \ln \left(\sqrt{8 \pi}\right) - \sqrt{\frac{\varpi}{12}}\ln \left( \lambda_0 \right),
\end{align}
and again represents singularity non-resolution. The variance of $\chi$ can be calculated as
\begin{align}
  \label{eq:71}
  \Delta_J \chi = \Braket{\chi^2}_J - \Braket{\chi}_J^2 = \frac{\varpi \left(T^2 + \gamma^2 \right)}{24 \gamma}.
\end{align}
$\Delta_J \chi$ is minimum when $\abs{T} \to 0$ and it diverges as
$\abs{T} \to \infty$.

With the parameter choices in~\cref{eq:69}, the expectation value of the Jordan
frame scale factor $a$ takes the form
\begin{align}
  \label{eq:74}
  &\Braket{a}_J = \Braket{\exp \left( \chi \right)}_J \nonumber \\
 = &\exp \left( \sqrt{\frac{P_T}{6}} \left( \sqrt{\varpi} - \sqrt{3} \right) T + \ln \left(\sqrt{8 \pi}\right) - \sqrt{\frac{\varpi}{12}}\ln \left( \lambda_0 \right) \right) \nonumber\\
  & \times \exp \left(  \frac{\varpi}{48} \left( \frac{T^2}{\gamma} + \gamma \right) \right)
\end{align}
which can be put in the form
\begin{align}
  \Braket{a}_J = \exp\left(\langle \chi \rangle_J+\frac{\Delta_J \chi}{2}\right) = a(T) \exp\left(\frac{\Delta_J \chi}{2}\right),
\end{align}
where $a(T)$ is the classical trajectory (from~\cref{eq:46})
\begin{align}
  \label{eq:75}
  &a(T) \nonumber\\
  &= \exp \left(  \sqrt{\frac{P_T}{6}} \left(  \sqrt{\varpi} - \sqrt{3} \right)T - \sqrt{\frac{\varpi}{12}} \ln \lambda_0 - \ln \sqrt{\frac{\kappa^2}{8 \pi}} \right) .
\end{align}
The relation between the expectation values of the scale factor and $\chi$ has a
similar structure to its Einstein frame counterpart~\cref{eq:58}. We see that
$\Braket{a}_J$ deviates from its classical trajectory, where the deviation is
parameterized by the fluctuations in $\chi$ and it increases as $\abs{T}$
becomes larger. Furthermore, the expectation value of scale factor represents a singularity free universe with its non-zero minimum being at $T=-24\sqrt{(P_T/6)}\gamma(\varpi-3)/\varpi$. The variance and relative fluctuation in the scale factor are
found to be
\begin{subequations}
  \begin{align}
    \label{eq:76}
    &\Delta_J a = \Braketil{a^2}_J - \Braket{a}_J^2 = \Braket{\exp \left( 2 \chi \right)}_J - \Braket{\exp \chi}_J^2\\
    &= \frac{8 \pi}{ \lambda_0^{ \sqrt{\frac{\varpi}{3}}}} \exp \left( 2  \sqrt{\frac{P_T}{6}} \left( \sqrt{\varpi} - \sqrt{3} \right) T +    2 \frac{\varpi}{48} \left( \frac{T^2}{\gamma} + \gamma \right)  \right) \nonumber\\
    &\times \left( \exp \left( \frac{\varpi \left( T^2 + \gamma^2 \right)}{24 \gamma} \right) - 1  \right),\\
    &    \frac{\sqrt{\Delta_J a}}{\Braket{a}_J} = \sqrt{\exp \left( \frac{\varpi \left(
      T^2 + \gamma^2 \right)}{24 \gamma} \right) - 1 }=\sqrt{\exp \left(
      \Delta_J\chi \right) - 1}.
  \end{align}
\end{subequations}
The relative fluctuation in the scale factor becomes minimum as $\abs{T} \to 0$
and it diverges in the limit $\abs{T} \to \infty$.

Starting with~\cref{eq:48}, we construct the operator for the Jordan frame
Hubble parameter as,
\begin{align}
  \label{eq:77}
  \hat{H} &= 16 \pi  \exp \left( T \sqrt{\frac{P_T}{2}} \right) \nonumber \\
          & \times \left( - \e^{-3 \hat \beta} \frac{1}{\varpi} \sqrt{\frac{P_T}{2}} - \frac{1}{2 \times 12} \left( \e^{- 3 \hat \beta} \hat P_\beta + \hat P _\beta  \e^{- 3 \hat \beta} \right)  \right).
\end{align}
The expectation of the Hubble parameter, with the parameter choice
from~\cref{eq:69}, becomes
\begin{subequations}
  \label{eq:78}
  \begin{align}
    \label{eq:79}
    \Braket{H}_J =&  \frac{\lambda_0 ^{\frac{\sqrt{3 \varpi}}{2}}}{2 \sqrt{3 \pi}} \frac{\sqrt{P_T}}{\varpi} \exp \left( T \sqrt{\frac{P_T}{2}} \left( 1 - \sqrt{3 \varpi} \right) \right) \nonumber \\
    & \times \exp \left( \frac{3 \varpi}{16} \left( \frac{T^2}{\gamma} + \gamma \right) \right) \left( \sqrt{\varpi} - \sqrt{3}  - \sqrt{\frac{6}{P_T}} \frac{\varpi}{8 \gamma} T \right)
    \intertext{which can be put in the form}
    \Braket{H}_J =&   H(T)  \exp \left(\frac{9}{2} \Delta_J \chi\right)\left( 1  - \sqrt{\frac{6}{P_T}} \frac{\varpi}{8 \gamma}  \frac{T}{\sqrt{\varpi}- \sqrt{3}}  \right),
  \end{align}
\end{subequations}
while, one can show that the classical expression for the Hubble parameter is
\begin{align}
  \label{eq:80}
  H (T) = \frac{\lambda_0 ^{\frac{\sqrt{3 \varpi}}{2}}}{2 \sqrt{3 \pi}} \frac{\sqrt{P_T}}{\varpi} \exp \left( T \sqrt{\frac{P_T}{2}} \left( 1 - \sqrt{3 \varpi} \right) \right) \left(\sqrt{\varpi} - \sqrt{3} \right).
\end{align}
Clearly, the expectation of the Hubble parameter differs from its classical
trajectory; the deviation increases as $\abs{T}$ becomes larger. In this case as well, the expectation value of the Hubble parameter represents a bouncing universe with bounce occurring at $T=8\sqrt{(P_T/6)}\gamma(\varpi-3)/\varpi$.
\section{Conformal map and expansion-collapse duality in the quantum description}
\label{sec:conf-map-expans}

We will now see how the classical conformal map translates into the quantum
descriptions of the Einstein and Jordan frame universes. In a previous
study~\cite{pandey.ea17}, the authors investigate the conformal map between the
Brans-Dicke Jordan frame and its corresponding Einstein frame in the quantum
description. It is argued that the classical conformal map holds true when
compared at the level of the Wheeler-DeWitt equations and the wave packets in
the two frames. However, the comparison at the level of wave packets may not be
sufficient, as the expectation values of different cosmological operators in the
two frames still may deviate from their classical conformal relations. In this
paper, we investigate the status of the conformal map in the quantum description
at the level of expectation values of relevant cosmological quantities.

In the previous section, the expectation values of different Einstein and Jordan
frame operators are obtained in terms of the clock variables in these frames,
$\tilde{T}$ and $T$, respectively. In order to compare the expectation values of
an operator in different frames, we need to consider the relation between
$\tilde{T}$ and $T$. In fact, one can use the conformal correspondence to obtain
the relation between the Einstein and Jordan frame clock variables as
\begin{align}
  \label{eq:81}
  \tilde{T} = \frac{\varpi}{2} \left( T - \frac{\ln \lambda_0}{\sqrt{2 P_T}} \right).
\end{align}
This shows that the clock variables are monotonically changing functions of each
other.

We have seen that classically, a Jordan frame with $\varpi<3$ contracts
indefinitely, corresponding to an expanding Einstein frame. In the quantum
description, we first find the relations between the expectation values of the
logarithmic scale factors, scale factors, and Hubble parameters in the two
frames, and compare these relations with their classical counterparts, set up by
the conformal transformation. Then we seek whether the quantum expectation
values of different operators suggest an expansion-collapse duality between the
frames, similar to the classical description. Finally, we find the relations
between quantum fluctuations in the two frames for different quantities. In
particular, we seek whether the rise in fluctuation in one frame implies the
same in the other, regardless of classical cosmological evolutions therein.

\subsection{Logarithmic scale factors}
\begin{figure}
  \centering
  \includegraphics[width=\columnwidth]{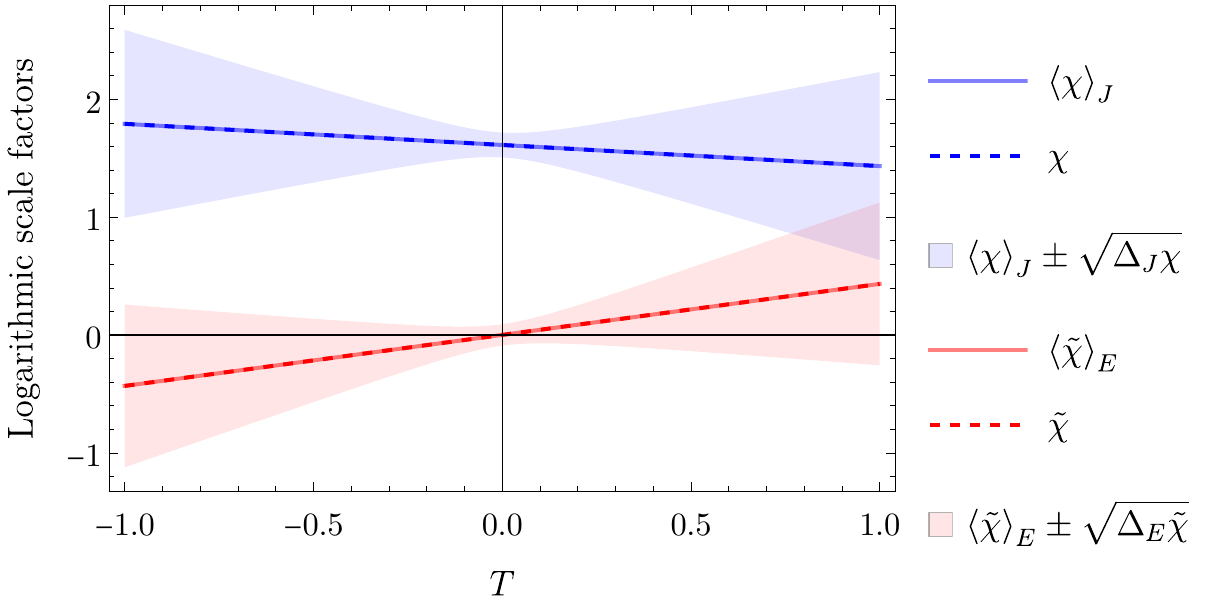}
  \caption{Evolution of the expectation values of the Jordan and Einstein frame
    logarithmic scale factors $(\Braket{\chi}_J, \Braket{\tilde{\chi}}_E)$
    (solid plots), along with their classical trajectories
    $(\chi, \tilde{\chi})$ (dashed plots), with respect to the Jordan frame
    clock $T$. The shaded regions depict fluctuations in the observables. We
    have used $\varpi=1.5,\ \gamma_E=0.1,\ \tilde{m}_0 = 1$, and other
    parameters are chosen according to~\cref{eq:53,eq:69,eq:84}. Both
    $\Braket{\chi}_J, \Braket{\tilde{\chi}}_E$ follow their corresponding
    classical trajectories throughout. The fluctuations in both the cases are
    minimum near $\abs{T} \to 0$ and get larger as $\abs{T}\to \infty $. Since
    $\varpi<3$, $\Braket{\chi}_J$ is always decreasing while $\Braket{\chi}_E$
    expands forever, consistent with the classical expansion-collapse duality
    scenario. }
  \label{fig:2}
\end{figure}

The classical conformal map between the Einstein and Jordan frame logarithmic
scale factors can be obtained from~\cref{eq:17} as
\begin{align}
  \label{eq:82}
  \tilde{\chi} - \chi =  \sqrt{\frac{P_T}{2}} T + \ln \frac{1}{\sqrt{8 \pi}},
\end{align}
where we have used the relations from \cref{eq:37} and \cref{eq:38}. On the other hand,
from \cref{eq:52,eq:68} we get
\begin{align}
  \label{eq:83}
  \langle \tilde{\chi} \rangle_E - \langle \chi \rangle_J = \sqrt{\frac{P_T}{2}} T + \ln \frac{1}{\sqrt{8 \pi}},
\end{align}
where we have replaced $\tilde{T}$ with $T$ using the conformal
relation~\cref{eq:81}, and the wave packet parameters are chosen according
to~\cref{eq:53,eq:69}. For these parameter choices, the RHSs of
the~\cref{eq:82,eq:83} are equal for all $T$ (see~\cref{fig:2}). Therefore, \emph{the expectation
  values of the Einstein and Jordan frame logarithmic scale factors follow the
  classical conformal relation for all time for a given choice of parameters}.

It is interesting to note that, given the parameter choices
from~\cref{eq:53,eq:69}, if we further impose
\begin{subequations}
  \label{eq:84}
  \begin{align}
    \label{eq:85}
    \tilde{\gamma} &= \frac{\varpi}{2}\gamma,
                     \intertext{and the integration constant}
                     \beta_0 &= 0 \implies \lambda_0 = 1,
  \end{align}
\end{subequations}
then the variances of $\tilde{\chi}$ and $\chi$ (from~\cref{eq:55,eq:71}) become
equal,
\begin{align}
  \label{eq:86}
  \Delta_E \tilde{\chi} = \Delta_J \chi.
\end{align}
Thus, \emph{for these parameter choices, the variance of logarithmic scale
  factor is conformally invariant.}

In order to see how the expectation values of the logarithmic scale factors
impose the condition for an expansion-collapse duality, let us consider the
following quantity
\begin{align}
  \label{eq:87}
  \diff{\Braket{\chi}_J}{\Braket{\tilde{\chi}}_E} = \frac{1}{\sqrt{\varpi}} (\sqrt{\varpi} - \sqrt{3}),
\end{align}
where we have used the parameter choices mentioned above. Given
$\Braket{\tilde{\chi}}_E$ is increasing with $T$, a Jordan frame with $\varpi<3$
will always have a decreasing $\Braket{\chi}_J$. The expansion-collapse duality
condition, in this case, is the same as the classical condition,~\cref{eq:18},
for all $T$. Therefore, in terms of the logarithmic scale factor operators, the
Jordan frame (with $\varpi<3$) is contracting indefinitely, corresponding to an
expanding Einstein frame, similar to the classical description. However, both
the expanding and contracting frames develop the same amount of quantum
fluctuation as $T$ increases, despite of their contrasting evolutions.

\subsection{Scale factors}
Classically, the conformal relation between the Einstein and Jordan frame scale
factors can be written from~\cref{eq:17} as
\begin{align}
  \label{eq:88}
  \frac{\at}{a} = \frac{1}{\sqrt{8 \pi}} \exp \left( \sqrt{\frac{P_T}{2}} T \right).
\end{align}
The ratio of the expectation values of the Einstein and Jordan frame scale
factors is obtain from~\cref{eq:57,eq:74} as
\begin{multline}
  \label{eq:89}
  \frac{\Braket{\at}_E}{\Braket{a}_J} = \frac{1}{\sqrt{8 \pi}} \lambda_0^{\frac{\varpi}{96 P_T \gamma} \left(\ln \lambda_0 - 2 \sqrt{2} \sqrt{P_T} T \right)} \\
  \exp \left( \sqrt{\frac{P_T}{2}} T + \frac{\varpi}{96} \left( \frac{\varpi}{\tilde{\gamma}} - \frac{2}{\gamma} \right) T^2  + \frac{1}{24} \left( \tilde{\gamma} - \frac{\varpi}{2} \gamma \right)\right),
\end{multline}
where we have imposed the parameters as in~\cref{eq:53,eq:69}, and replaced
$\tilde{T}$ with $T$ from~\cref{eq:81}. If we further consider the parameter
choices that ensure conformally invariant variance of $\chi$, from~\cref{eq:84},
it takes the form
\begin{align}
  \label{eq:90}
  \frac{\Braket{\at}_E}{\Braket{a}_J} = \frac{1}{\sqrt{8 \pi}} \exp \left( \sqrt{\frac{P_T}{2}}T \right),
\end{align}
the RHS of which is the same as that of~\cref{eq:88}. Therefore, though the
expectation values of the Einstein and Jordan frame scale factors do not follow
their corresponding classical trajectories for all $T$, the ratio of them is the
same as the ratio of their classical counterparts, provided the choice of
parameters. \emph{The expectations of the scale factors of Einstein and Jordan
  frames satisfy the classical conformal transformation relation for a certain
  choice of wave packet parameters and boundary condition.}

For the same parameter choice from~\cref{eq:53,eq:69,eq:84}, the quantum
fluctuations of the Einstein and Jordan frame scale factors are related as
\begin{align}
  \label{eq:91}
  \sqrt{\frac{\Delta_E \at}{\Delta_J a}} &= \frac{1}{\sqrt{8 \pi}} \exp \left( \sqrt{\frac{P_T}{2}} T \right) = \frac{\Braket{\at}_E}{\Braket{a}_J}. 
\end{align}
That is, the ratio of the roots of variances of Einstein and Jordan frame scale
factors is the same as the ratio of the expectations of the scale factors, which
is according to the classical conformal map. Hence, \emph{the relative
  fluctuation of scale factors is invariant under the conformal transformation,}
\begin{align}
  \label{eq:92}
  \frac{\sqrt{\Delta_E \at}}{\Braket{\at}_E} =  \frac {\sqrt{\Delta_J a}}{\Braket{a}_J}. 
\end{align}

\begin{figure}
  \centering
  \includegraphics[width=\columnwidth]{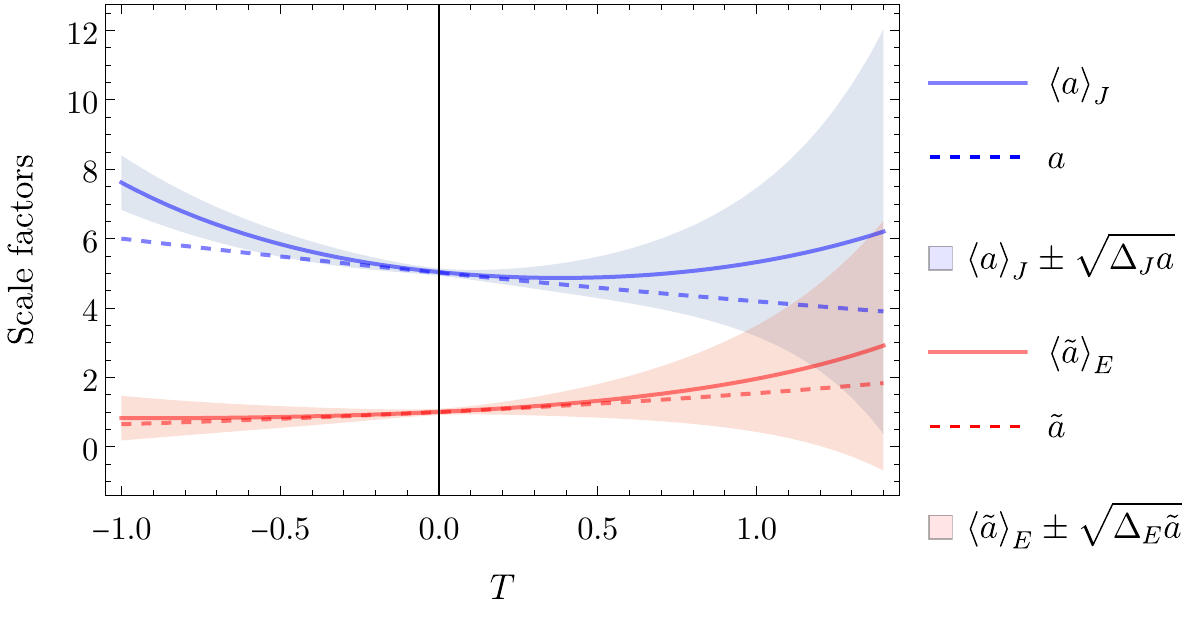}
  \caption{Evolution of the expectation values of the Jordan and Einstein frame
    scale factors $(\Braket{a}_J, \Braket{\tilde{a}}_E)$ (solid plots), along
    with their classical trajectories $(a, \tilde{a})$ (dashed plots), with
    respect to the Jordan frame clock $T$. The shaded regions depict
    fluctuations in the observables. We have used
    $\varpi=1.5,\ \gamma_E=0.1,\ \tilde{m}_0 = 1$, and other parameters
    according to~\cref{eq:53,eq:69,eq:84}. Both
    $\Braket{a}_J, \Braket{\tilde{a}}_E$ follow their corresponding classical
    trajectories near $\abs{T} \to 0$, and deviate as $\abs{T}$ gets larger. The
    fluctuations in both the cases also get larger as $\abs{T}\to \infty $.
    Although $\Braket{a}_J$ is decreasing near $\abs{T} \to 0$, it eventually
    starts to increase at a larger $T$, while the classical trajectory $a(T)$
    decreases forever, since $\varpi<3$. In terms of
    $\Braket{a}_J, \Braket{\tilde{a}}_E$, the classical expansion-collapse
    scenario for $\varpi<3$ is only observed near $\abs{T} \to 0$.}

  \label{fig:3}
\end{figure}

The condition for the expansion-collapse duality, as imposed by the scale factor
operators, can be written as
\begin{subequations}
  \label{eq:93}
  \begin{align}
    \label{eq:94}
    \diff{{\Braket{a}_J}}{{\Braket{\at}_E}} &< 0,
                                            \intertext{where}
                                            \diff{{\Braket{a}_J}}{{\Braket{\at}_E}} &= \diff{{\Braket{a}_J}}{T} \diff{T}{{\Braket{\at}_E}} = \diff{{\Braket{a}_J}}{T} \diff{{\tilde T}}{{\Braket{\at}_E}} \diff{T}{\tilde{T}},\\
    \diff{{\Braket{a}_J}}{{\Braket{\at}_E}} &= \frac{2\sqrt{\pi}}{\sqrt{\varpi}} \frac{24 \sqrt{P_T} \gamma \left( \sqrt{\frac{\varpi}{3}} - 1 \right) + \sqrt{2}\varpi T}{4 \sqrt{6 P_T} \gamma + \sqrt{\varpi} T}.
  \end{align}
\end{subequations}
In the limit $\abs{T} \to \infty$,
$\frac{\d {\Braket{a}_J}}{\d {\Braket{\at}_E}} > 0$, the scale factors'
expectation values in the two frames change monotonically with each other
regardless of $\varpi$. Therefore, according to the scale factor operators,
there is no possibility of an expansion-collapse duality as $\abs{T} \to \infty$
for all $\varpi$. This is in contrast with the classical condition as well as
the condition set by the logarithmic scale factor operators. However, the
quantum fluctuations in both the scale factors are large in this limit. In the
$\abs{T} \to 0$ limit, when the quantum fluctuations are small,
$\frac{\d {\Braket{a}_J}}{\d {\Braket{\at}_E}} < 0$ requires $\varpi < 3$,
indicating that the classical condition for the expansion-collapse duality is
recovered in this limit (see~\cref{fig:3}).

\subsection{Hubble parameters}
Let us now consider the relation between the expectation values of the Hubble
parameters in the two frames. Classically, the ratio of the Jordan and Einstein
frame Hubble parameters can be obtained using the conformal correspondence as
\begin{align}
  \label{eq:95}
  \frac{H}{\tilde{H}} &= \frac{1}{\sqrt{8 \pi}} \exp \left( \sqrt{\frac{P_T}{2}} \right) \left( 1 - \sqrt{\frac{3}{\varpi}} \right).
\end{align}
As it is expected, given $\tilde{H}>0$, the Jordan frame Hubble parameter $H$
becomes negative for $\varpi<3$, depicting a contracting Jordan frame. With the
parameter choices from~\cref{eq:53,eq:69,eq:84}, and using the relation
in~\cref{eq:81}, one can write the ratio of the expectations of Jordan and
Einstein frame Hubble parameters (\cref{eq:63,eq:79}) as
\begin{subequations}
  \label{eq:96}
  \begin{align}
    \label{eq:97}
    \frac{\Braket{H}_J}{\Braketil{\tilde{H}}_E} = \exp \left( \sqrt{\frac{P_T}{2}T} \right) \frac{24 \sqrt{P_T} \gamma \left(1 - \frac{\sqrt\varpi}{\sqrt 3} \right) + 3 \sqrt{2} \varpi T}{4 \sqrt{\pi} \left( 3 \varpi T - 4 \sqrt{6} \gamma \sqrt{\varpi P_T} \right)}.
  \end{align}
\end{subequations}
\begin{figure}
  \centering
  \includegraphics[width=\columnwidth]{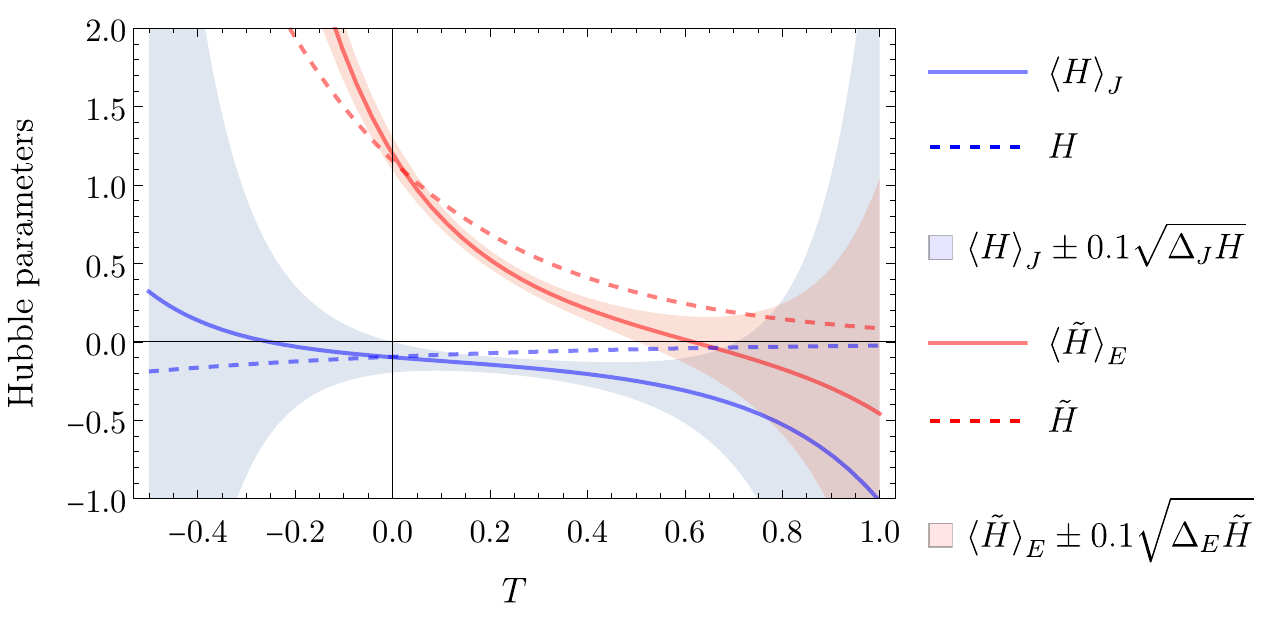}
  \caption{Evolution of the expectation values of the Jordan and Einstein frame
    Hubble parameters $(\Braketil{H}_J, \Braketil{\tilde{H}}_E)$ (solid plots),
    along with their classical trajectories $(H, \tilde{H})$ (dashed plots),
    with respect to the Jordan frame clock $T$. The shaded regions depict
    fluctuations in the observables. We have used
    $\varpi=1.5,\ \gamma_E=0.1,\ \tilde{m}_0 = 2$, and other parameters
    according to~\cref{eq:53,eq:69,eq:84}. Both
    $\Braketil{H}_J, \Braketil{\tilde{H}}_E$ align with their corresponding
    classical trajectories only near $\abs{T} \to 0$, and deviate as $\abs{T}$
    gets larger. The fluctuations in both the cases also get larger as
    $\abs{T}\to \infty $. Near $\abs{T} \to 0$, $\Braketil{H}_J<0$ and
    $\Braketil{\tilde{H}}_E>0$, therefore indicating an expansion-collapse
    duality similar to their classical counterparts. However, at a larger $T$,
    both $\Braketil{H}_J, \Braketil{\tilde{H}}_E < 0$, indicating both the
    frames are contracting; while at a smaller negative $T$
    $\Braketil{H}_J, \Braketil{\tilde{H}}_E > 0$, indicating both the frames are
    expanding. In terms of $\Braketil{H}_J, \Braketil{\tilde{H}}_E$, the
    classical expansion-collapse scenario for $\varpi<3$ is observed near
    $\abs{T} \to 0$.}
  \label{fig:4}
\end{figure}
Clearly, the ratio deviates from its classical counterpart \cref{eq:95}, which
is due to the classical conformal map. In particular, in the limit
$\abs{T} \to \infty$,
$\frac{\langle H \rangle_J}{\langle \tilde{H} \rangle_E} > 0$ regardless of the
choice of $\varpi$; hence there is no expansion-collapse duality in this limit,
as determined by the expectation values of the Hubble parameters. However, the
quantum fluctuations in both the Einstein and Jordan frame Hubble operators are
large at this regime (see~\cref{fig:4}). The expansion-collapse requirement set
by the Hubble operators in this regard is similar to that of the scale factor
operators, which is in contrast with the classical condition as well as the
condition set by the logarithmic scale factor operators. The quantum
fluctuations in both the Hubble parameters become small as $\abs{T} \to 0$. In
this limit, $\frac{\langle H \rangle_J}{\langle \tilde{H} \rangle_E} $ becomes
negative for $\varpi<3$, thus the classical condition for the expansion-collapse
duality is recovered in this limit~(see~\cref{fig:4}).

\section{Summary and conclusion}
\label{sec:summary-conclusion}

We study the conformal correspondence between Jordan and Einstein frames in a
quantum mechanical framework. In particular, we consider a Brans-Dicke model
without a potential in the Jordan frame, which corresponds to a massless scalar
field in the Einstein frame. According to the classical conformal map, the
Jordan frame universe may contract indefinitely to an arbitrarily small scale
factor, while the Einstein frame continually expands. When the scale factor in
the contracting Jordan frame becomes sufficiently small, and the universe
approaches the singularity, the classical description of the system becomes
inadequate. One expects the Jordan frame universe to develop significant quantum
characteristics in this regime, for example, rise in quantum fluctuations in
different cosmological quantities. However, the corresponding Einstein frame
universe is arbitrarily large at this point, therefore it is expected to behave
classically. This leads to the apparent paradox -- does the conformal
correspondence in this scenario provide a map between a quantum effect-dominated
universe and a universe behaving classically?

To consistently address this question, we quantize the Einstein and
Jordan frame universes following the Wheeler-DeWitt prescription. We first
investigate whether the classical conformal map is valid at the quantum level by
comparing the expectation values of different cosmological
quantities in the two frames. We then find how the quantum fluctuations in
different quantities are related in the two frames. The main results in this
work can be summarized as follows:
\begin{itemize}
\item For a convenient choice of phase-space variables, the Wheeler-DeWitt
  equations in both frames take the simple form of the Schr\"odinger equation
  for a free particle. The wave packets constructed in both the Einstein and
  Jordan frames satisfy DeWitt's criteria, and therefore hint at singularity
  avoidance. However, the expectation values of both Einstein and Jordan frame
  logarithmic scale factors follow their corresponding classical trajectories
  for all values of the clock parameter $(T)$. In this case, following their
  classical trajectories, the expectation values do not lead to singularity
  avoidance. On the other hand, the expectation values of the scale factors and
  Hubble parameters in both frames deviate from their classical trajectories as
  $\abs{T}$ becomes larger. Eventually, the expectation values of these
  operators represent singularity-free bouncing universes.
  
\item \emph{The expectation values of the logarithmic scale factors in the two
    frames always satisfy the classical conformal correspondence for a given set
    of parameter choices}. Moreover, for the same set of parameters, \emph{the
    variance in the logarithmic scale factor operator is found to be conformally
    invariant}, it takes the same value in both Einstein and Jordan frames at a
  given clock parameter ($T$). The variance is minimum as $\abs{T} \to 0$, and
  it diverges as $\abs{T} \to \infty$.

\item Similarly, \emph{the expectation values of the scale factors in the two
    frames also satisfy the classical conformal map}, even though the
  expectation values in this case do not always follow the classical
  trajectories. \emph{The relative fluctuation in the scale factor operator is
    also conformally invariant}. In both the frames, the variances in the scale
  factors are minimum near $\abs{T} \to 0$, when the expectations value of the
  scale factors follow the classical behavior. They keep on increasing as
  $\abs{T}$ increases, and the expectation values keep on deviating from the
  classical trajectories.

\item Unlike the logarithmic scale factors and the scale factors, the
  expectation values of the Hubble parameters do not always follow the classical
  conformal relation. However, similar to the other two operators, the quantum
  variances in both the Einstein and Jordan frame Hubble parameters are minimum
  in the $\abs{T} \to 0$ limit, and they keep on increasing as $\abs{T}$
  increases. Similar to the scale factor operators, the expectation values of
  the Hubble parameters in both frames deviate from the classical trajectories
  as $\abs{T}$ becomes larger.

\item In terms of the logarithmic scale factor operators, the Einstein and
  Jordan frames undergo indefinite expansion and contraction, respectively, set
  by the classical condition of the expansion-collapse duality. However,
  according to both the scale factors and Hubble operators, there is no eternal
  expansion-collapse duality between the frames, but only in the limit when
  $\abs{T} \to 0$, and the quantum fluctuations in these operators are small.
  This contrasts with the classical duality, as well as the expansion-collapse
  duality dictated by the logarithmic scale factors.
\end{itemize}
It is evident that different cosmological operators may disagree on whether one
frame is expanding or contracting as $\abs{T}$ becomes large and the
fluctuations in the operators increase. As $\abs{T} \to 0$, the fluctuations in
the operators decrease, and their expectation values resemble their
corresponding classical trajectories. In this limit, the expectation values of
all the three operators depict an expansion-collapse duality between the
conformal frames, similar to the classical description.

With regard to the initial question, we find that the classical conformal map
holds true at the quantum level when compared through the expectation values of
the logarithmic scale factors and the scale factors in the two frames. As the
Jordan frame contracts and its classical scale factor approaches singularity in
the limit $T \to \infty$, the quantum fluctuations in different cosmological
operators expectedly become large, indicating the rise in the quantum
characteristics in the Jordan frame universe. More interestingly, at the same
time, the quantum fluctuations also increase in the expanding Einstein frame,
despite its arbitrarily large scale factor. \emph{Therefore, the rise in quantum
  fluctuations in one frame indicates the same in the other, even though the
  cosmological evolutions in the frames are drastically different.}

The present results indicate that the rise in quantum features is a frame
independent effect, even the expanding conformal universe with arbitrarily large
size can harbor large quantum fluctuations. The presence of significant quantum
effects in a large expanding universe is previously argued in the literature.
For example, \cite{alexandre2022} predicts non-trivial quantum effects when the
physical universe transitions from the decelerating phase to an accelerating
one. In \cite{dhanuka2020}, it is argued that quantum fluctuations may remain
non-vanishing in the late-time universe and can become dominant over large
scales. The revival of quantum correlations at the late-time matter-dominated
universe is explored in~\cite{dhanuka2022}.

This resolution of the apparent paradox at hand leads to another paradoxical
notion of large quantum fluctuations in a macroscopic universe. One can naively
expect the decoherence mechanism to suppress large quantum effects in the
background geometry of the macroscopic universe. However, the universe is an
isolated system, and there is no notion of an external environment to decohere
the system. Still, in canonical quantum gravity, decoherence can be achieved by
considering the inhomogeneous degrees of freedom, such as density fluctuations
and quasi-normal modes, as the environment, while the background degrees of
freedom as the system~\cite{kiefer2012,Kiefer_2003,Zeh_1986}. For example,
inhomogeneous modes are found to decohere the quantized background geometry for
WKB-like solutions of Wheeler-DeWitt equation
\cite{Kiefer_1992,Barvinsky_1999,Demers_1996,Kiefer_2001}. However, a similar
notion of a decohering environment does not exist for the model under
consideration; therefore, it is not surprising to find large quantum
fluctuations for the macroscopic universe in this case.

In this paper, the quantum analysis is done with convenient choices of clock
variables, phase-space variables, and operator ordering of the Hamiltonians. It
is worth exploring whether different choices of clock variables influence the
results~\cite{gielen2020,gielen2022}. The present work can be extended by
introducing ordinary matter components in the two conformal frames. It is argued
in~\cite{banerjee2016} that in the presence of matter, the conformal
correspondence breaks down at the level of the wave packets. It would be
interesting to investigate the conformal map in such a scenario at the level of
expectation values of relevant cosmological operators. This will be pursued in a
future work.


\section*{Acknowledgement}
\label{sec:ack}

The authors thank Kinjalk Lochan for his important comments and suggestions
during the preparation of the manuscript and for careful reading of the
manuscript. DM thanks Narayan Banerjee for valuable discussion on the topic. The
authors would also like to thank the anonymous referee for their insightful
comments on a previous draft of the manuscript. Research of DM is partially
supported by the Department of Science and Technology, India, through a project
grant under DST/INSPIRE/04/2016/000571. HSS would like to acknowledge the
financial support from the University Grants Commission, Government of India, in
the form of Junior Research Fellowship (UGC-CSIR JRF/Dec- 2016/503905).


\input{draft.bbl}
\end{document}

%% file: settings.tex
\usepackage{braket}
\usepackage{graphicx}
\usepackage{braket}
\usepackage{url}
\usepackage{float}
\usepackage[thinc]{esdiff}
\usepackage[dvipsnames]{xcolor}
\usepackage{tensor}
\usepackage{esdiff}
\usepackage{amsmath, amssymb}
\usepackage[pdfusetitle,bookmarksnumbered=true]{hyperref}
\usepackage{cleveref}
\usepackage{empheq}
\usepackage{subcaption}
\usepackage[most]{tcolorbox}
\hypersetup{colorlinks = true,
  linkcolor = blue,
  anchorcolor = blue,
  citecolor = gray,
  filecolor = blue,
  urlcolor = blue }
\usepackage{commath}
\usepackage{tensor}
\usepackage{tocloft}
\usepackage[font={small}]{caption}
\usepackage{authblk}

\newcommand{\p}{\partial}

\newcommand{\at}{\tilde{a}}
\renewcommand{\d}{\text{d}}
\newcommand{\te}{\tilde{t}}
\newcommand{\e}{\text{e}}
\newcommand{\wbd}{w_{\text{BD}}}
\renewcommand{\H}{\mathbb{H}}
\newcommand{\Te}{\tilde{T}}

\renewcommand{\abs}[1]{|#1|}
\newcommand{\Braketil}[1]{\langle#1\rangle}